\documentclass[12pt]{article}
\usepackage{cite}
\newcommand{\fnd}[2]{\frac{\textstyle #1}{\textstyle #2}}
\newcommand{\xrm}[1]{{\textstyle \mbox{\rm #1}}}
\newcommand{\bm}[1]{\mbox{\boldmath $#1$}}
\newcommand{\abs}[1]{\left| #1\right|}
\newcommand{\bracket}[2]{\mbox{$\left\langle #1\left| #2\right.\right
\rangle$}}
\newcommand{\slsh}[1]{{\not\!{#1}}}

\newcommand{\Real}[1]{\Re {\it e}\left( #1 \right)}
\newcommand{\Imag}[1]{\Im {\it m}\left( #1 \right)}
\begin{document} \baselineskip .7cm
\title{$W$-boson unitarity effects and
oscillations in the neutral Kaon system}
\author{
Eef van Beveren\\
{\normalsize\it Centro de F\'{\i}sica Te\'{o}rica}\\
{\normalsize\it Departamento de F\'{\i}sica, Universidade de Coimbra}\\
{\normalsize\it P-3000 Coimbra, Portugal}\\
{\small email: eef@teor.fis.uc.pt\hspace{0.3cm}
URL: http://cft.fis.uc.pt/eef}\\ [.3cm]
\and
George Rupp\\
{\normalsize\it Centro de F\'{\i}sica das Interac\c{c}\~{o}es Fundamentais}\\
{\normalsize\it Instituto Superior T\'{e}cnico, Edif\'{\i}cio Ci\^{e}ncia}\\
{\normalsize\it P-1049-001 Lisboa Codex, Portugal}\\
{\small email: george@ajax.ist.utl.pt}\\ [.3cm]
{\small PACS number(s): 11.30.Er, 13.25.Es, 13.38.Be, 11.10.St}\\ [.3cm]
{\small hep-ph/0308030}
}
\date{\today}
\maketitle

\begin{abstract}
We study the effects on oscillations in the neutral-Kaon system
from lepton and hadron self-energy loops in the $W$-boson propagators. The
corresponding $W$ box diagrams are evaluated by attaching the external quark
lines to covariant vertex functions for the composite Kaons, integrating
over all off-shell momenta. We find that the ratio of the imaginary and real
parts of the amplitude is of the same size as the experimental value
of the modulus of the $CP$-violating parameter $\varepsilon$.
\end{abstract}

The theory of weak decays, initially formulated by Fermi
\cite{NC11p1,ZP88p161} in terms of a four-fermion vertex for the
description of neutron $\beta$ decay, is nowadays understood by a
renormalizable \cite{NPB7p637,NPB35p167} non-abelian gauge theory
\cite{NP22p579,PRL19p1264,PL13p168},
in which the weak interactions are mediated by the heavy gauge bosons.
Here we shall concentrate on $K^0$-$\bar{K}^0$ oscillations via
double $W$ exchanges.

The phenomenon of
a long-lived component in the two-pion decay mode of
the neutral Kaon system,
discovered by Christenson, Cronin, Fitch, and Turlay \cite{PRL13p138}, is in
the Standard Model (SM) parametrized by the complex phase of the
Cabibbo\--Kobayashi\--Maskawa (CKM) \cite{PRL10p531,PTP49p652} matrix,
which is related to the complex $\varepsilon$ parameter.
This $\varepsilon$ measures the fraction
of pion pairs observed in
the decay products of the long-lived neutral Kaons
\cite{MRR,ISMP103}.
Many authors have studied this phenomenon, see e.g.\
Refs.~\cite{PL15p1965,PR9p143,NPB247p70,PRL55p1039,BLOIS89p3,N344p197,
FPL11p303,HEPPH9807514,HEPPH9811508,EPJC13p267,
AJP69p44,HEPPH0004245,HEPPH0201064,
HEPPH0205164,HEPPH0207082,IJMPA18p1551,HEPPH0307119}, just to mention a few.
But its dynamics remains to be discovered.

The masses of the heavy gauge bosons, which can be estimated from Fermi's
weak coupling constant, are experimentally determined with great
precision \cite{PRD66p010001}, ever since their first discovery
in 1983 by the UA2 collaboration \cite{PLB122p476}.
Furthermore, the branching ratios of the $W$s and $Z$ are well known
\cite{PRD66p010001}
for most of the decay modes that could play a role at Kaon energies.

In scattering theory, unstable particles are de\-scrib\-ed by Breit-Wigner
amplitudes \cite{PR49p519}, associated with complex-energy poles in the
$S$ matrix for the decay products \cite{HEPPH0304105}.
Consequently, the physical, on-shell $W$ boson, which besides being massive
(mass $M_{W}$), is also unstable (width $\Gamma_{W}$),
should be characterized by the complex quantity
$M_{W}-\frac{1}{2}i\Gamma_{W}$.

In this paper, instead of describing the oscillations in
the neutral Kaon system by a complex CKM phase
and the exchange of massive $W$ bosons with purely real mass $M_{W}$,
we study whether a comparable
result may be obtained by considering only real
CKM matrix elements \cite{PRD65p075014}, but allowing the exchanged off-shell
$W$ bosons to develop a complex self-energy whenever this is kinematically
demanded by four-momentum conservation and the known decay thresholds of the
$W$. The crucial point here is that \em ``decay is a profoundly irreversible
process'' \em \cite{Bohm2003}, which should also be taken into account in
virtual exchanges, especially when hunting after a tiny effect. In order to
have a reliable off-shell calculation, we shall sandwich the standard box
diagrams between covariant composite-Kaon vertex functions that
are parameter-free, after being tuned to the experimental Kaon size.
Note that the Kaon's compositeness has recently been suggested to
even give rise to $CP$ violation \cite{HEPPH0211005}.

\begin{figure}[htbp]
\begin{center}
\begin{picture}(350,107)(0,0)
\put(90,78){\vector(1,0){10}}
\put(180,78){\vector(1,0){10}}
\put(250,78){\vector(1,0){10}}
\put(90,74){\makebox(0,0)[tc]{$d$}}
\put(180,74){\makebox(0,0)[tc]{$u$, $c$, $t$}}
\put(270,74){\makebox(0,0)[tc]{$s$}}
\put(110,82){\makebox(0,0)[bc]{$p_{1}$}}
\put(180,82){\makebox(0,0)[bc]{${p_{1}}'$}}
\put(250,82){\makebox(0,0)[bc]{${p_{1}}''$}}
\put(110,22){\vector(-1,0){10}}
\put(260,22){\vector(-1,0){10}}
\put(90,27){\makebox(0,0)[bc]{$\bar{s}$}}
\put(180,25){\makebox(0,0)[bc]{$\bar{u}$, $\bar{c}$, $\bar{t}$}}
\put(270,27){\makebox(0,0)[bc]{$\bar{d}$}}
\put(110,15){\makebox(0,0)[tc]{$p_{2}$}}
\put(180,18){\makebox(0,0)[tc]{${p_{2}}'$}}
\put(250,18){\makebox(0,0)[tc]{${p_{2}}''$}}
\put(50,50){\makebox(0,0){\large \bm{K^{0}}}}
\put(310,50){\makebox(0,0){\large \bm{\bar{K}^{0}}}}
\put(135,50){\makebox(0,0)[rc]{\large $W$}}
\put(225,50){\makebox(0,0)[lc]{\large $W$}}
\put(145,50){\makebox(0,0)[lc]{$q_{1}$}}
\put(215,50){\makebox(0,0)[rc]{$q_{2}$}}
\put(50,96){\makebox(0,0)[bc]{\bm{\Gamma}}}
\put(310,96){\makebox(0,0)[bc]{\bm{\bar{\Gamma}}}}
\end{picture}
\end{center}
\caption[]{Diagram $T$: The $t$-channel box diagram for
$K^{0}$ - $\bar{K}^{0}$ oscillations.}
\label{boxt}
\end{figure}
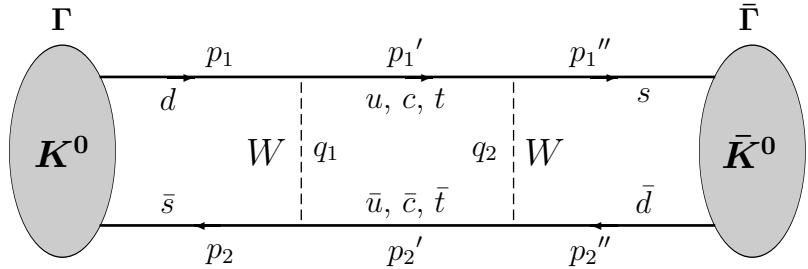

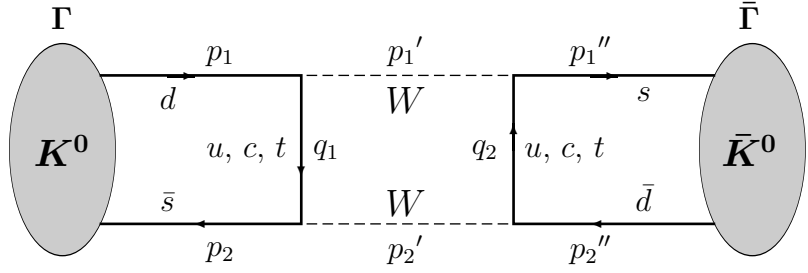
\begin{figure}[htbp]
\begin{center}
\begin{picture}(350,120)(0,0)
\put(90,78){\vector(1,0){10}}
\put(140.5,50){\vector(0,-1){10}}
\put(110,22){\vector(-1,0){10}}
\put(90,74){\makebox(0,0)[tc]{$d$}}
\put(180,74){\makebox(0,0)[tc]{\large $W$}}
\put(270,74){\makebox(0,0)[tc]{$s$}}
\put(110,82){\makebox(0,0)[bc]{$p_{1}$}}
\put(180,82){\makebox(0,0)[bc]{${p_{1}}'$}}
\put(250,82){\makebox(0,0)[bc]{${p_{1}}''$}}
\put(260,22){\vector(-1,0){10}}
\put(220.8,50){\vector(0,1){10}}
\put(250,78){\vector(1,0){10}}
\put(90,27){\makebox(0,0)[bc]{$\bar{s}$}}
\put(180,25){\makebox(0,0)[bc]{\large $W$}}
\put(270,27){\makebox(0,0)[bc]{$\bar{d}$}}
\put(110,15){\makebox(0,0)[tc]{$p_{2}$}}
\put(180,18){\makebox(0,0)[tc]{${p_{2}}'$}}
\put(250,18){\makebox(0,0)[tc]{${p_{2}}''$}}
\put(50,50){\makebox(0,0){\large \bm{K^{0}}}}
\put(310,50){\makebox(0,0){\large \bm{\bar{K}^{0}}}}
\put(135,50){\makebox(0,0)[rc]{$u$, $c$, $t$}}
\put(145,50){\makebox(0,0)[lc]{$q_{1}$}}
\put(225,50){\makebox(0,0)[lc]{$u$, $c$, $t$}}
\put(215,50){\makebox(0,0)[rc]{$q_{2}$}}
\put(50,96){\makebox(0,0)[bc]{\bm{\Gamma}}}
\put(310,96){\makebox(0,0)[bc]{\bm{\bar{\Gamma}}}}
\end{picture}
\end{center}
\caption[]{Diagram $S$: The $s$-channel box diagram for
$K^{0}$ - $\bar{K}^{0}$ oscillations.}
\label{boxs}
\end{figure}

In principle, one could calculate the desired amplitude by directly
determining the two-pion branching ratio of the long-lived neutral Kaon
\cite{IJTP36p2239}. However, such a calculation would involve the unknows of
QCD at low energies, which causes the result to depend on the specific
modeling of strong interactions.
Here, we determine the lowest-order processes that
transform neutral Kaons into neutral anti-Kaons, schematically represented
by the diagrams $T$ and $S$ of Figs.~\ref{boxt} and \ref{boxs}, respectively,
and compare our result with the modulus of
$\varepsilon$. From experiment \cite{PLB420p191} one derives
$\abs{\varepsilon}=0.0023\pm 0.0002$.

The strategy to obtain the modulus of $\varepsilon$
from a microscopic description stems from Gaillard and Lee
\cite{PRD10p897}. It is based on a conjecture of Wolfenstein \cite{PRL13p352}
that all $CP$-violating processes stem from double-strangeness-changing
($\Delta S=2$) superweak interactions.
The transition matrix elements for neutral-Kaon oscillations
are related to the modulus of $\varepsilon$ by
\begin{equation}
\abs{\varepsilon}\; =\;\frac{1}{2}\fnd
{\bracket{K^{0}}{\bar{K}^{0}}-\bracket{\bar{K}^{0}}{K^{0}}}
{\bracket{K^{0}}{\bar{K}^{0}}+\bracket{\bar{K}^{0}}{K^{0}}}
\;\;\; ,
\label{abseps}
\end{equation}
which equals half the ratio of the imaginary and real parts of the amplitude.

The original calculus \cite{PRD10p897} was carried out by taking the external
quarks free and the external momenta zero.
However, we know that free quarks do not exist. Moreover, the neutral Kaons
have long lifetimes --- on a hadronic scale --- thus justifying  a bound-state
approach with respect to their wave functions from strong interactions.
Finally, the intermediate state (e.g.\ $u\bar{u}$) may have a Kaon-like
distribution w.r.t.\ strong interactions, but the involved $u$ (or $c$, $t$)
and $\bar{u}$ (or $\bar{c}$, $\bar{t}$) quarks can have all possible momenta,
only governed by total four-momentum conservation. This picture is to be
contrasted with the on-shell $\Delta I\!=\!1/2$ $K_S\to\pi^+\pi^-$ decay via
an intermediate $\sigma$ resonance \cite{PRD53p2421}.
Hence, also for the intermediate state one must expect rearrangement effects
in the quark-antiquark strong-interaction distributions.
Therefore, we shall take the external quark momenta of the box
diagram in agreement with the bound-state picture for neutral Kaons.
As a consequence, our quarks will be dressed, and so massive.
We take for the constituent quark masses the reasonable values
$m_{n}=m_{u}=m_{d}=340$ MeV and $m_{s}=490$ MeV \cite{IJMPA13p657}.
In any case, we shall see later that the results hardly depend on these
specific values.
The initial and final quark-antiquark bound states couple, through the
vertex functions, to the quark and antiquark participating
in the $W$-boson exchanges.
The precise details of the vertex functions must be deduced from
a microscopic model.
We assume here that the vertex functions $\Gamma$ and $\bar{\Gamma}$
depend on the relative quark-antiquark four-momenta
$p$ and $p''$, respectively, and on the total center-of-mass (CM) momenta
$P=p_{1}+p_{2}$ resp.\ $P''={p_{1}}''+{p_{2}}''$, i.e.,
\begin{equation}
\Gamma\; =\;\Gamma (p,P)
\;\;\;\;\xrm{and}\;\;\;\;
\bar{\Gamma}\; =\;\bar{\Gamma}(p'',P''\,)
\;\;\; .
\label{qqbarvertex}
\end{equation}
Total four-momentum conservation is expressed by
\begin{equation}
P\; =\; P'\; =\; {p_{1}}'+{p_{2}}'\; =\; P''
\;\;\; .
\label{Econserve}
\end{equation}
For the $K^{0}$ and $\bar{K}^{0}$ vertex functions
$\Gamma$ resp.\ $\bar{\Gamma}$ we assume a Euclidean
Gaussian distribution ($\exp\left[ -\frac{1}{2}\alpha p^{2}\right]$), with
$\alpha =5.37$ GeV$^{-2}$, which in the nonrelativistic limit corresponds to
the experimental Kaon charge radius of 2.84 GeV$^{-1}$ \cite{PRD66p010001}, and
is, moreover, in good agreement with the predictions of the quark-level linear
$\sigma$ model \cite{NPA724p391}
and also with the wave functions obtained in a
unitarized meson model \cite{ZPC19p275}.

The relations between the particle momenta $p_{1}$ and $p_{2}$
and the relative particle momenta $p$ are given by the standard
expressions
\begin{equation}
p_{1,2}\; =\;\pm p\; +\;\fnd{m_{n,s}}{m_{n}+m_{s}}\; P
\;\;\; ,
\label{relmomentum}
\end{equation}
and similar relations for ${p_{1}}'$, ${p_{2}}'$ and $p'$, and
${p_{1}}''$, ${p_{2}}''$ and $p''$.
The definitions (\ref{relmomentum}) have a nonrelativistic origin, but,
as our calculation is fully covariant, the final result does not
depend on this particular choice, which is nonetheless very convenient for
numerical reasons.

For the $W$-propagators we write $S^{(i)}_{W}(q_{i},M_{W})$, where
the exchange momenta $q_{1}$ and $q_{2}$ are defined in Fig.~\ref{boxt},
according to $q_{1}={p_{1}}'-{p_{1}}$ and $q_{2}={p_{1}}''-{p_{1}}'$.
The two-particle propagators $G_{q\bar{q}}(p,P)$ have the usual form
\begin{equation}
G_{d\bar{s}}(p,P)\; =\; \fnd{1}
{\left( p_{1}^{2}-m_{n}^{2}+i\epsilon\right)
\left( p_{2}^{2}-m_{s}^{2}+i\epsilon\right)}
\;\;\; ,
\label{2pp}
\end{equation}
and similarly for the other two.
Now, we are dealing here with the propagators of quarks, which are
fermions. Hence the two-particle propagators should be of the form
\begin{equation}
G\left( p_{1},p_{2}\right)\; =\;\fnd{1}
{\left(\slsh{p_{1}}-m_{1}+i\epsilon\right)
\left(\slsh{p_{2}}-m_{2}+i\epsilon\right)}
\;\;\; .
\label{2ppfermion}
\end{equation}
This can be arranged by multiplying the two-particle propagators of
Eq.~(\ref{2pp}) by
\begin{equation}
\left(\slsh{p_{1}}+m_{1}\right)\left(\slsh{p_{2}}+m_{2}\right)
\;\;\; .
\label{2ppfermion1}
\end{equation}
However, for the crucial off-shell kinematical effects only the pole
structure is of importance, which is the same for the fermion progators
(\ref{2ppfermion}) and the boson propagators (\ref{2pp}). Then, all the
factors resulting from the Dirac algebra involving expressions like in
Eq.~(\ref{2ppfermion1}), and also the different SM vertices, will
cancel out when evaluating the ratio of the imaginary and real parts of the
amplitude. Note that working with only scalar propagators allows us to use
scalar vertex functions for the Kaons, and thus to model the diagrams in an
essentially parameter-free way.

The total amplitude for the process represented by diagram $T$ is given by
\begin{equation}
A\propto\int d^{4}p\, d^{4}p'\, d^{4}p''\,
\Gamma\, G_{d\bar{s}}\, S^{(1)}_{W}\, G_{u\bar{u}}\, S^{(2)}_{W}\,
G_{s\bar{d}}\,\bar{\Gamma} .
\label{texchange}
\end{equation}
The integrations are performed numerically, except for some angular
integrations.

The pole structure of the two-particle propagators can best be studied in
the CM frame ($\sqrt{s}$ represents the total invariant mass).
If we then define
\begin{equation}
\omega_{n,s}\; =\;\sqrt{{\vec{p}\;}^{2}+m_{n,s}^{2}}
\;\;\; ,
\end{equation}
we obtain for the terms in the denominator of
the two-particle propagator (\ref{2pp}) the expressions
\begin{eqnarray}
p_{1}^{2}-m_{n}^{2}+i\epsilon & = &
\left(\fnd{m_{n}}{m_{n}+m_{s}}\sqrt{s}+p_{0}\right)^{2}-
\omega_{n}^{2}+i\epsilon \;\;\;,
\nonumber\\
& & \label{2pp1}\\ [-.3cm]
p_{2}^{2}-m_{s}^{2}+i\epsilon & = &
\left(\fnd{m_{s}}{m_{n}+m_{s}}\sqrt{s}-p_{0}\right)^{2}-
\omega_{s}^{2}+i\epsilon
\;\;\; .
\nonumber
\end{eqnarray}
We have poles in the complex $p_{0}$-plane for the two-particle propagator
(\ref{2pp}) at
\begin{eqnarray}
& (1,2):\; &
p_{0}\; =\; -\fnd{m_{n}}{m_{n}+m_{s}}\sqrt{s}\pm
\left(\omega_{n}^{2}-i\epsilon\right)\;\;\;\xrm{and}
\nonumber\\
& (3,4):\; &
p_{0}\; =\; \fnd{m_{s}}{m_{n}+m_{s}}\sqrt{s}\pm
\left(\omega_{s}^{2}-i\epsilon\right)
\;\;\; ,
\label{2pp1poles}
\end{eqnarray}
graphically represented in Fig.~(\ref{polestructure}).
\begin{figure}[htbp]
\begin{center}
\begin{picture}(230,60)(0,25)
\put(215,53){\makebox(0,0)[bc]{\bm{\Real{p_{0}}}}}
\put(118,75){\makebox(0,0)[lc]{\bm{\Imag{p_{0}}}}}
\put(20,60){\makebox(0,0)[bc]{\bm{\ast}}}
\put(90,60){\makebox(0,0)[bc]{\bm{\ast}}}
\put(140,40){\makebox(0,0)[tc]{\bm{\ast}}}
\put(210,40){\makebox(0,0)[tc]{\bm{\ast}}}
\put(20,70){\makebox(0,0)[bc]{\bm{(2)}}}
\put(90,70){\makebox(0,0)[bc]{\bm{(4)}}}
\put(140,30){\makebox(0,0)[tc]{\bm{(1)}}}
\put(210,30){\makebox(0,0)[tc]{\bm{(3)}}}
\end{picture}
\end{center}
\caption[]{Singularity structure of formula (\ref{2pp1}).
The poles ($\ast$) are numbered according to their appearance in
Eq.~(\ref{2pp1poles}).}
\label{polestructure}
\end{figure}

Notice that $\sqrt{s}<m_{n}+m_{s}$ in the bound-state regime.
Therefore, the Wick rotation $p_{0}\rightarrow ip_{4}$, and similarly for
$p'$ and $p''$, can be freely carried out (see e.g.\ Ref.~\cite{PRC37p1729}),
resulting in a purely real amplitude,
at least when neglecting the $W$ self-energy effects.
\clearpage

For the full determination of $\bracket{K^{0}}{\bar{K}^{0}}$, we also
include the $s$-channel double-$W$-exchange diagram depicted in
Fig.~\ref{boxs}.
We obtain contributions for diagram $S$ which in magnitude
are comparable to those for diagram $T$.
Moreover, diagram $S$ requires less numerical effort than
diagram $T$, since in the former case all angular integrations
can be performed analytically. Consequently, checking details on how our
results depend on $\alpha$ and the quark masses we only do for diagram $S$.

\begin{figure}[htbp]
\begin{center}
\begin{picture}(100,50)(0,-15)
\put(65,10){\makebox(0,0)[bl]{\bm{W}}}
\end{picture}
\begin{picture}(20,30)(0,-15)
\put(10,-0.8){\makebox(0,0){\bm{=}}}
\end{picture}
\begin{picture}(60,30)(0,-15)
\end{picture}
\begin{picture}(20,30)(0,-15)
\put(10,0){\makebox(0,0){\bm{+}}}
\end{picture}
\begin{picture}(100,30)(0,-15)
\put(50,15){\makebox(0,0)[bc]{\bm{e, \mu, \tau}}}
\put(50,-15){\makebox(0,0)[tc]{\bm{\nu_{e}, \nu_{\mu}, \nu_{\tau}}}}
\end{picture}
\begin{picture}(20,30)(0,-15)
\put(10,0){\makebox(0,0){\bm{+}}}
\end{picture}
\begin{picture}(20,30)(0,-15)
\put(10,0){\makebox(0,0){\bm{\cdots}}}
\end{picture}
\end{center}
\caption[]{The full W propagator dressed with lepton loops and possibly
with hadron loops}
\label{dressedW}
\end{figure}
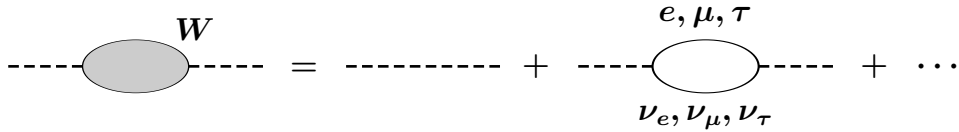

Now, let us finally come to the main issue of this paper.
In order to treat the exchanged $W$ resonances in a more realistic way than
what is normally done, we first include self-energy bubbles \cite{HEPPH9903219}
corresponding to the well-known \cite{PRD66p010001} leptonic decay modes of the
$W$, i.e., $e\nu_{e}$, $\mu\nu_{\mu}$, and $\tau\nu_{\tau}$,
as depicted in Fig.~\ref{dressedW}.

Hence, we substitute $M_{W}^{2}$ in the $W$ propagators by
$M_{W}^{2}+Z_{e}+Z_{\mu}+Z_{\tau}$,
according to
\begin{equation}
Z_{\ell}\; =\;
-\frac{i}{2}\;\fnd{\Gamma_{\ell}}{\abs{p_{\ell}}}\; M_{W}^{2}\;
\left( 1-\fnd{m_{\ell}^{2}}{p^{2}}\right) \;
\Theta\!\left(1-\fnd{m_{\ell}^2}{p_{\mbox{\scriptsize Mink.}}^2}\right)
\;\;\; ,
\label{Wimag}
\end{equation}
where $\Gamma_{\ell}$, $\abs{p_{\ell}}$, $m_{\ell}$ ($\ell=e$, $\mu$, $\tau$)
and $M_{W}$ are taken from Ref.~\cite{PRD66p010001}. Note that in the $\Theta$
function one must take the Minkowskian $p^2=p_0^2-\vec{p}^{\,2}$, with $p_0$
real, too. For obvious reasons, we have only kept the imaginary parts of the
contributions to the mass self-energy (Fig.~\ref{dressedW}), the real parts
being absorbed in the physical $W$ central mass. At this point, we should
also stress that adding a negative imaginary part (\ref{Wimag}) to the $W$
mass does \em not \em \/spoil our straightforward Wick rotations, as it only
amounts to substituting an infinitesimal $\epsilon$ by a finite $|Z(p^2)|$. In
other words, because of causality the sign of the imaginary part of the mass is
always negative, both for stable and unstable (anti)particles
\cite{AIP660p325}.

In Table~\ref{leptonsandhadrons} we show how much each of the terms $Z_{\ell}$
contributes to the ratio in Eq.~(\ref{abseps}).

\begin{table}[htbp]
\begin{center}
\begin{tabular}{|l||c|c|}
\hline & & \\ [-.3cm]
mode & fraction \cite{PRD66p010001} &
$\fnd{\Imag{\bracket{K^{0}}{\bar{K}^{0}}}}
{\Real{\bracket{K^{0}}{\bar{K}^{0}}}}$\\ [-.3cm]
& & \\
\hline & & \\ [-.4cm]
$e\nu_{e}$ & (10.72$\pm$0.16)\% & 0.00096\\
$\mu\nu_{\mu}$ & (10.57$\pm$0.22)\% & 0.00094\\
$\tau\nu_{\tau}$ & (10.74$\pm$0.27)\% & 0.00060\\
$c\bar{s}$ ($D\bar{K}$) & ($31^{+13}_{-11}$)\% & 0.00153\\
$cX$ ($D\bar{B}$) & (33.6$\pm$2.7)\% & 0.00078\\ [-.4cm]
& & \\
\hline
\end{tabular}
\end{center}
\caption[]{Contributions to the ratio of the imaginary part over the real part
of the $K^0$-$\bar{K}^0$ amplitude from the most important $W$ decay modes
(see  also text).}
\label{leptonsandhadrons}
\end{table}

For the hadronic modes we make the following considerations.
The only known mode containing a light hadron is $\pi\gamma$,
having a decay rate of less than 0.1\% \cite{PRD66p010001}, hence negligible.
The next level of known modes contains at least one $c$ quark.
The threshold for the $DK$ loops at 2.37 GeV is some 600 MeV higher than
for $\tau\nu_{\tau}$, whereas the branching ratio is about the triple of the
latter, which yields a contribution about two and a half times larger than
from $\tau\nu_{\tau}$ (see Table~\ref{leptonsandhadrons}).

The fraction of $cX$ decay modes of the $W$ is a bit larger than three times
the fraction of $\tau\nu_{\tau}$.
However, here we do not have much information on $X$.
If we assume $X=b$, we get a contribution from $DB$ loops
which is smaller than from $DK$, but still considerable.

It is clear from these considerations that we cannot include the hadronic
loops to a great accuracy, because of the experimental uncertainties.
Nevertheless, we may conclude from our results that their total contribution
will not be very different from the sum of the two individual cases
included by us, which is a kind of upper bound.
Nevertheless, we shall take a conservative choice for the total final
theoretical error in our result.
Summing the contributions of Table~\ref{leptonsandhadrons}
(we actually put all loops together when determining the total
$\bracket{K^{0}}{\bar{K}^{0}}$ amplitude), we obtain
\begin{equation}
\fnd{\Imag{\bracket{K^{0}}{\bar{K}^{0}}}}
{2\Real{\bracket{K^{0}}{\bar{K}^{0}}}}\; =\;
{0.0024\;}^{+0.0002}_{-0.0004}
\;\;\;
\label{lepsandhads}
\end{equation}
for the total contribution of $W$ self-energy loops,
which is of the same magnitude as $\abs{\varepsilon}$ \cite{PRD66p010001}.
In the errors quoted in Eq.~(\ref{lepsandhads}), we have included, besides
small numerical errors, the effects of reasonable variations in the parameters
$\alpha$, $m_ {n}$, and $m_ {s}$, as well as the uncertainty in the $cX$ mode.
Also the contributions for intermediate $c$ and $t$ quarks have been studied,
which do not significantly alter the result (\ref{lepsandhads}), though adding
to the total $\bracket{K^{0}}{\bar{K}^{0}}$ amplitude.
The coincidence of the effect computed by us and $\abs{\varepsilon}$ is
very striking, the physical significance of which is open to debate.
\vspace{.3cm}

{\large\bf Acknowledgments}
\vspace{.3cm}

We are indebted to M.~K.~Gaillard for very pertinent criticism
concerning the first version of this paper.
We thank M.~D.~Scadron and F.~Kleefeld for useful discussions as well as
critical comments.
This work is partly supported by the
{\it Funda\c{c}\~{a}o para a Ci\^{e}ncia e a Tecnologia}
of the {\it Minist\'{e}rio da
Ci\^{e}ncia e do Ensino Superior} \/of Portugal,
under contract numbers POCTI/\-35304/\-FIS/\-2000 and
POCTI/\-FNU/\-49555/\-2002.

\end{document}